\documentclass[a4paper,12pt]{article}
\usepackage{graphicx}
\usepackage{amsmath,amssymb}

\def\beqra{\begin{eqnarray}}
\def\eeqra{\end{eqnarray}}
\def\beq{\begin{equation}}
\def\eeq{\end{equation}}

\begin{document}
\baselineskip=15.5pt
\pagestyle{plain}
\setcounter{page}{1} 

\setlength\arraycolsep{2pt}

\begin{titlepage}


\rightline{\small{DESY 07-145}}

\begin{center}

\vskip 1.7 cm

{\LARGE {\bf Axion--Dilaton Cosmology\\ and Dark Energy}}

\vskip 1.5cm
{\large 
Riccardo Catena and Jan M\"oller  
}
\vskip 1.2cm

{\it Deutsches Elektronen-Synchrotron DESY, Theory Group, \\
Notkestrasse 85, D-22603 Hamburg, Germany}

\vskip 0.5cm

\vspace{1cm}

{\bf Abstract}

\end{center}

We discuss a class of flat FRW cosmological models based on D=4 axion-dilaton gravity universally coupled to cosmological background fluids. In particular, we investigate the possibility of recurrent acceleration, which was recently shown to be generically realized in a wide class of axion-dilaton models, but in absence of cosmological background fluids. We observe that, once we impose the existence of radiation --~and~matter~-- dominated earlier stages of cosmic evolution, the axion-dilaton dynamics is altered significantly with respect to the case of pure axion-dilaton gravity. During the matter dominated epoch the scalar fields remain either frozen, due to the large expansion rate, or enter a cosmological scaling regime. In both cases, oscillations of the effective equation of state around the acceleration boundary value are impossible. Models which enter an oscillatory stage in the low redshift regime, on the other hand, are disfavored by observations. We also comment on the viability of the axion-dilaton system as a candidate for dynamical dark energy. In a certain subclass of models, an intermediate scaling regime is succeeded by eternal acceleration. We also briefly discuss the issue of dependence on initial conditions.

\noindent

\end{titlepage}

\newpage

\section{Introduction}

Although no fundamental scalar particles have been discovered yet, many attempts
to extend the Standard Model of particle physics naturally introduce new
scalar and/or pseudo--scalar degrees of freedom, {\it e.g.}
scalar superpartners of Standard Model fields in a supersymmetric framework, 
moduli fields related to geometric properties of compactified extra dimensions, {\it etc}.
Whereas on the one hand this leads to challenging problems \cite{moduliproblem}, 
on the other hand these new fields could provide
interesting candidates of dynamical dark energy \cite{DEreview}.

Of particular interest in cosmology are low-energy effective theories where the gravitational sector includes, apart from the metric tensor, scalar (or pseudo-scalar) degrees of freedom that could provide a {\it gravitational interpretation} of early (inflationary), or recent (quintessential), periods of acceleration. 

In the extensively studied case of Scalar--Tensor (ST) theories of gravity \cite{ST}, the existence of 
one or more scalar partners of the graviton leads to modifications of the Hubble expansion
and of Newton's law \cite{ST1, Damour}. While ultra-light scalar fields
\footnote{Exhibiting a mass of the order of the present value of the Hubble parameter}
are in general potentially dangerous sources of new long range forces, interestingly enough, ST theories 
are protected against any violation of the weak equivalence principle by a universal metric
coupling between matter and the gravity sector \cite{Damour}. Because of this property, ST theories
provide a natural framework to address the issue of dynamical dark energy \cite{STDE}. 
On the other hand, inflation can also be successfully achieved in a ST picture \cite{inflation}.

The ST scenario can be generalized to include also pseudo-scalar fields. This is the case, for instance, of axion--dilaton (AD) gravity, which can be viewed as a prototype of theories where a ``dilaton-like'' scalar and an ``axion-like'' pseudo-scalar appear as spin zero partners of the graviton. Such a picture naturally emerges from the Neveu-Schwarz bosonic sector of the low-energy string effective action \cite{Witten}. Black hole solutions of this theory have been found in \cite{blackhole}, while domain wall solutions were given in \cite{domainswall}.
A more general class of stationary supersymmetric solutions was discussed in \cite{Kallosh}. Cosmological implications of such a theory -- up to first order in perturbation theory -- were investigated in \cite{Copeland}, where for the first time the spectrum of cosmological perturbations was computed. 

It has been pointed out recently that AD gravity theories with an exponential dilaton potential admit cosmological solutions which give rise to the interesting phenomenon of recurrent acceleration \cite{Sonner}. By a detailed phase-space analysis of the AD dynamical system, the authors verified the generic occurrence of recurrent acceleration in the regime of a spiral focus associated to a runaway behavior of both fields.
In this picture, the present acceleration does not appear as a peculiar stage of the cosmic history, being likely a transient or even recurring phenomenon. In particular, they conclude that the future evolution of the universe is by no means determined to be accelerating forever, in obvious contrast to standard $\Lambda$CDM cosmology. 

However, in order to relate these results to the observed accelerated expansion of the universe, it is crucial to take into account the non-gravitational sector of the theory. While Sonner and Townsend \cite{Sonner} considered models comprising only the graviton, the axion and the dilaton fields, the purpose of the present paper is to investigate how the contribution of cosmological perfect fluids -- (dark) matter and radiation -- modifies the evolution of the AD system. Avoiding any attempt of constructing a fully realistic model, we assume that the gravity sector fields couple in a universal metric way to the background fluid. This choice is inspired by the aforementioned ST theories.

The main result of our analysis is that recurrent acceleration is no longer a generic feature of the (modified) AD dynamical system, once we impose the pre-existence of a radiation (RDE) and a matter dominated (MDE) era and take into account the finite contribution of (dark) matter to the present energy density.

Our paper is organized as follows. In section \ref{background} we present the cosmological evolution equations of AD gravity coupled to matter in a universal metric way. Due to this coupling any matter field experiences gravitational interactions through the same metric $\tilde{g}_{\mu \nu}$, which is conformally related to the Einstein frame metric by a dilaton-dependent function. As it will be made explicit in section \ref{background}, the dilatonic part of the interaction is parametrized by a function $Q$. The simplest case of minimal coupling corresponds to $Q=0$ and will be discussed separately in sections \ref{nofluid} and \ref{Q0}; the more general case of constant $Q > 0$ is treated in section \ref{Q}. In section \ref{Conclusions}, which is devoted to our conclusions, we also briefly comment on the viability of AD models as candidates of dynamical dark energy. In the appendices we summarize some basic facts concerning dynamical systems terminology and give details of the calculations.

\newpage
\section{Axion--dilaton cosmology}
\label{background}    

The class of models we are interested in is described by the following action
\beq
S = S_{\textrm{AD}} + S_{\textrm{fluid}}\,,
\label{action}
\eeq
where
\beq
S_{\textrm{AD}}= \int d^4x \sqrt{-g} 
\Bigg\{ \frac{1}{2}R- \frac{1}{2}\partial_{\mu}\Phi\partial^{\mu}\Phi 
- \frac{1}{2}e^{-\gamma\Phi} \partial_{\mu}\sigma\partial^{\mu}\sigma - e^{-\lambda\Phi} \Bigg\}\notag
\eeq
and
\beq
S_{\textrm{fluid}} = S_{\textrm{fluid}}[A^{2}(\Phi)g_{\mu\nu},\Psi] \,.\notag
\eeq
In Eq.(\ref{action}) $R$ is the Ricci scalar constructed from the Einstein 
frame metric $g_{\mu\nu}$, $A(\Phi)$ is an arbitrary function of the dilaton 
field (to be specified below) and $\gamma,$ as well as $\lambda>0,$ are real constant parameters. 
The background fluid sector is described by the action $S_{\textrm{fluid}}$.
Hereafter ``background fluid sector'' refers to the sector of the theory that includes all the fields $\Psi$
of the Standard Model (or of one of its possible extensions)
which we assume to be coupled to the gravitational sector ($S_{\textrm{AD}}$) by means of the same metric 
$\tilde{g}_{\mu\nu}~=~A^{2}(\Phi)~g_{\mu\nu}$.

Strictly speaking, by this choice we neglect any non-universal couplings of the AD system
, in particular interactions with the field strength 
of some gauge field, {\it i.e.} $\sim \Phi\,F_{\mu\nu} F^{\mu\nu}$ in the case of the dilaton, 
and $\sim \sigma\, F_{\mu\nu} \tilde{F}^{\mu\nu}$ in the case of the axion. 
In fact, such couplings are generically present in any theory which couples AD gravity 
to a matter and gauge sector (see for example \cite{Dilaton, Dilrun}). 
However, a proper treatment of these interactions and their consequences in a cosmological framework
is beyond the scope of this paper. On the other hand, our analysis applies -- more generally -- to 
any complex scalar field with modular invariant 
\footnote{Invariant under  $SL(2,\mathbb{R})-$transformations
$\tau \to \frac{a \tau + b}{c \tau + d}$ with $a d - b c=1$ where 
$\tau = \frac{\gamma}{2} \sigma + i e^{-\frac{\gamma}{2} \Phi}$.}
kinetic term. 

The introduction of an exponential potential for the dilaton explicitly breaks 
the $SL(2,\mathbb{R})$ invariance of the usual AD gravity.
Such a potential with $2 \lambda = \gamma$ emerges, for instance, from 
a truncation of the Freedman--Schwarz $ D = 4$ supergravity theory \cite{Sugra} 
(see also section $4.3$ of \cite{domainswall} for an alternative motivation).    

In a flat FRW Universe, 
\beq
ds^2 = -dt^2 + a^{2}(t)d\vec{x}^2 \,,
\eeq
the cosmological equations obtained from the action (\ref{action}) are
\newpage
\begin{align}
\ddot{\Phi} &= -3H\dot{\Phi} + \lambda e^{-\lambda \Phi} - \frac{1}{2}\gamma e ^{-\gamma \Phi}\dot{\sigma}^2 
+Q(\Phi)\rho_{\textrm{fluid}}\,(1-3 w_{\textrm{fluid}})\,, \label{eomd}\\
\ddot{\sigma} &= -(3H-\gamma\dot{\Phi})\,\dot{\sigma}\,, \label{eoma}\\ 
\dot{H} &=-\frac{1}{2}\left[ (1 + w_{\textrm{fluid}})\rho_{\textrm{fluid}} + 
\dot{\Phi}^2 + e ^{-\gamma \Phi} \dot{\sigma}^2 \right]\,,\\
H^2 &=\frac{1}{3}
\left[ \rho_{\textrm{fluid}} + \frac{1}{2}( \dot{\Phi}^2 
+ e ^{-\gamma \Phi}\dot{\sigma}^2) + e^{-\lambda\Phi} \right] \,\label{system},
\end{align}
where $Q(\Phi)$ is given by 
\beq
Q(\Phi) \equiv -\frac{d \ln A}{d\Phi} \,.
\eeq
The background fluid energy density $\rho_{\textrm{fluid}}$ obeys the continuity equation
\beq
\dot{\rho}_{\textrm{fluid}}= \left[-3(1+w_{\textrm{fluid}})\,H-(1-3w_{\textrm{fluid}})\,Q(\Phi)\,\dot{\Phi}
\right] \rho_{\textrm{fluid}},
\eeq
with equation of state parameter $w_{\textrm{fluid}} \in [0,\frac{1}{3}],$ the limit values corresponding 
to pure matter and pure radiation respectively. 

Introducing the following dynamical variables,
\beq
x_1^2 \equiv \frac{\dot{\Phi}^2}{6H^2}\,, \quad x_2^2 \equiv \frac{e^{-\gamma\Phi}\dot{\sigma}^2}{6H^2}\,,\quad
y^2 \equiv \frac{V(\Phi)}{3H^2}\,, \quad z^2 \equiv \frac{\rho_{\textrm{rad}}}{3H^2} \,,
\eeq
the system~(\ref{eomd})--(\ref{system}) can be rewritten in autonomous form, straightforwardly generalizing the set--up of \cite{CLW},
\begin{align}
\frac{dx_1}{dN}&=\frac{3}{2}\,x_1(x_1^2+x_2^2-y^2+\frac{1}{3}z^2-1)
+\sqrt{\frac{3}{2}}\left[-\gamma x_2^2+\lambda y^2
+Q(1-x_1^2-x_2^2-y^2-z^2)\right],\label{eqn:DSi}\\
\frac{dx_2}{dN}&=
\frac{3}{2}\,x_2(x_1^2+x_2^2-y^2+\frac{1}{3}z^2-1)
+\sqrt{\frac{3}{2}}\gamma\, x_1 x_2,\\
\frac{dy}{dN}&=\frac{3}{2}\,y\,(x_1^2+x_2^2-y^2+\frac{1}{3}z^2+1)-\sqrt{\frac{3}{2}}\lambda\, x_1 y,\\
\frac{dz}{dN}&=\frac{3}{2}\,z\,(x_1^2+x_2^2-y^2+\frac{1}{3}(z^2-1)).\label{eqn:DSf}
\end{align}
where $N=\ln a$, and $Q$ is assumed to be a positive real constant, corresponding to $A(\Phi)\sim \exp(-Q\Phi)$.

The given system of evolution equations defines a three-parameter family of dynamical models with four-dimensional 
compact phase-space,
$$x_1^2+x_2^2+y^2+z^2\leq 1,$$
and can, furthermore, be restricted to $(x_2,y,z)\geq 0$, 
since the system is invariant under change of sign in any of these variables. 

Using the new set of variables, the effective equation of state parameter can be conveniently expressed as $$w_{\textrm{eff}}\equiv
\frac{p_{\textrm{AD}}+p_{\textrm{fluid}}}{\rho_{\textrm{AD}}+\rho_{\textrm{fluid}}}
=x_1^2+x_2^2-y^2+\frac{1}{3}z^2\,.$$
The equation $w_{\textrm{eff}}(x_1, x_2, y, z)=-1/3$ defines the boundary of the domain of accelerated expansion in phase-space.

The purpose of the succeeding section is to reproduce the findings of \cite{Sonner} within the chosen framework, 
corresponding to the theory of Eq.(\ref{action}) 
truncated by $S_{\textrm{fluid}}=0$.

\section{Axion-dilaton dynamics ($S_{\textrm{fluid}}= 0$)}
\label{nofluid}

Provided a flat FRW universe and absence of a cosmological perfect fluid, 
the phase-space of the AD system is two-dimensional. 
We choose it to be spanned by $\lbrace x_1, x_2\rbrace.$ 
The Friedmann constraint equation now reads $$x_1^2+x_2^2+y^2=1,$$ 
and we can eliminate $y$ from the system:
\beqra
\frac{dx_1}{dN}&=& 3\,x_1(x_1^2+x_2^2-1)
+\sqrt{\frac{3}{2}}[-\gamma x_2^2+\lambda(1-x_1^2-x_2^2)] \,, \label{eq1} \\
\frac{dx_2}{dN}&=& 3\,x_2(x_1^2+x_2^2-1)
+\sqrt{\frac{3}{2}}\gamma\, x_1 x_2 \,.
\label{eq2}
\eeqra

The equation of state is then given by
$$w_{\textrm{eff}}=w_{\textrm{AD}}=2(x_1^2+x_2^2)-1.$$
Each model is characterized by a number of stationary solutions, or critical points, of the corresponding autonomous system. (We refer to the appendix concerning a brief summary of relevant terminology.) We find the following set of stationary points $X_s=(x_{1,s},x_{2,s})$, given as functions of the parameters:

\begin{align}
B_1,B_2:\quad &(\pm 1 ,0),\notag\\
G:\quad &\left(\frac{\lambda}{\sqrt{6}} ,0\right),\notag\\
J:\quad &\left(\frac{\sqrt{6}}{\gamma+\lambda} ,
\sqrt{\frac{\lambda(\gamma+\lambda)-6}{(\gamma+\lambda)^2}}\right).\notag
\end{align}

\begin{table}[h]\begin{center}
\begin{tabular}{|c|c|c|c|}
\hline
\textbf{fixed point}&\textbf{existence}&\textbf{stability}
&\textbf{$w$}\\
\hline
\hline
$B_1$ & $\forall (\gamma,\lambda)$ & stable: $\gamma <0 \wedge \lambda > \sqrt{6}$ & 1\\
$B_2$ & $\forall (\gamma,\lambda)$ & saddle point: $\gamma >0$ & 1\\
\hline
$G$ & $\lambda < \sqrt{6}$ & stable:  $\lambda(\lambda+\gamma)<6$ & $-1+\frac{\lambda^2}{3}$\\[2pt]\hline
$J$ & $\gamma\geq 0 ~\wedge ~\lambda(\lambda+\gamma)\geq 6$ & stable: $\gamma>0 ~\wedge ~\lambda(\lambda+\gamma)> 6$& $\frac{\lambda-\gamma}{\lambda+\gamma}$\\[2pt]
\hline
\end{tabular}
\caption{Properties of the fixed points of the reduced dynamical system.}
\label{tab:SonTown}
\end{center}\end{table}

Properties of the fixed points are displayed in table \ref{tab:SonTown}. 
The existence condition can be expressed as follows,
$$x_1^2+x_2^2\leq 1,$$
with $x_1,x_2$ real. The stability of a critical point is determined by the eigenvalues of the Jacobi matrix
$$M:=\left(\frac{\partial F_i}{\partial x_j}\right)_{i,j\,\epsilon\,\lbrace 1,2\rbrace}$$
of the vector function
\begin{align}
F(x_1, x_2)=\Bigg\{ &3\,x_1(x_1^2+x_2^2-1)+\sqrt{\frac{3}{2}}(-\gamma x_2^2+\lambda(1-x_1^2-x_2^2),\notag\\
&3\,x_2(x_1^2+x_2^2-1)+\sqrt{\frac{3}{2}}\gamma\, x_1 x_2 \Bigg\},
\end{align}
evaluated at $X_s$. (See the appendix for more details.)
We find the following eigenvalues:

$$B_1:\quad\sqrt{\frac{3}{2}}\gamma,\quad 6-\sqrt{6}\lambda,$$

$$B_2:\quad -\sqrt{\frac{3}{2}}\gamma,\quad 6+\sqrt{6}\lambda,$$

$$G:\quad\frac{1}{2}(\lambda^2 -6),\quad \frac{1}{2}(\lambda(\lambda+\gamma)-6),$$

$$J:\quad \frac{3}{2(\gamma+\lambda)}\left(-\gamma \pm
\sqrt{\gamma^2+8\gamma(\gamma+\lambda)-\frac{4}{3}\gamma\lambda(\gamma+\lambda)^2}\right).$$\\

The fixed point $J$ is a spiral focus if
\beq
3\gamma\,(9\gamma+8\lambda)-4\,\gamma\lambda(\gamma+\lambda)^2<0.
\label{eq:spiral}
\eeq

\subsection{Recurrent acceleration}
\label{Recacc}

We now discuss under which conditions recurrent periods of acceleration can
be realized within the family of dynamical models given by Eqs.~(\ref{eq1})~--~(\ref{eq2}).
 
Let us first observe that,
according to table \ref{tab:SonTown},
accelerated expansion is possible at the fixed point $G$ -- if $\lambda<\sqrt{2}$ -- 
or at the fixed point $J$ -- if $\gamma>2\lambda$. Then, restricting ourselves to the case $\gamma>0$, 
we can distinguish three possibilities to realize a model which generically allows for periods 
of accelerated expansion. We give examples of phase portraits of the different cases below.

As we will see, in agreement with \cite{Sonner}, recurrent periods of acceleration are efficiently
produced by models corresponding to a subset of parameter space where the fixed point $J$ 
is stable and a spiral focus (figure \ref{STparam}). 

\subsubsection{$G$ stable}

If $\lambda<\sqrt{2},$ the attractor $G$ is situated within the domain of accelerated expansion. In this case, once acceleration has set in, it will last forever 
. If $\sqrt{2}<\lambda<\sqrt{6},$ it can be a transient phenomenon along a subset of trajectories (figure \ref{fig:phaseplot(8/5,1)}). 

\begin{figure}[t]
\begin{center}
\includegraphics[width=10cm]{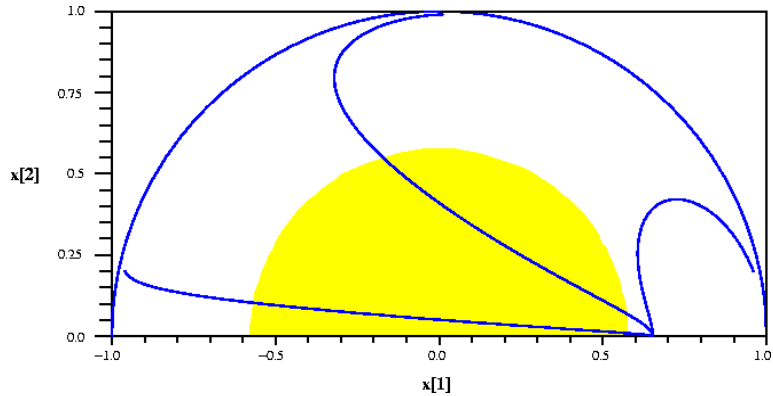}
\caption{Model with parameters $(\lambda,\gamma)=\left(\frac{8}{5},1\right)$.Shaded (yellow) area corresponds to accelerated expansion. Trajectories are plotted in the plane $(x_{1},x_{2})$.}
\label{fig:phaseplot(8/5,1)}
\end{center}
\end{figure}

\subsubsection{$J$ stable, $G$ saddle point}
\label{STlim}

If $J$ is the attractor, the domain of acceleration in parameter space is bounded by $\gamma=2\lambda$. The phase-portrait of the system depends crucially on the progress of the special trajectory connecting the saddle point $G$ with the attractor. We will hereafter call it the {\it connecting trajectory}.
If $\lambda<\sqrt{2}$, both $G$ and $J$ are situated within the domain of acceleration in phase-space, and hence the connecting trajectory is completely contained within this domain as well. Any trajectory approaching the connecting one will therefore remain inside the acceleration domain once having entered it (see figure \ref{fig:phaseplot(sqrt{2},4)}). 
If, on the other hand, $G$ is situated outside, recurrent acceleration can be generically realized, {\it if} the spiral focus $J$ is located close enough to the acceleration boundary, such that any trajectory approaching the attractor crosses the boundary repeatedly, as does the connecting trajectory (figure \ref{fig:phaseplot(2,4)}).

\begin{figure}[t]
\begin{center}
\includegraphics[width=10cm]{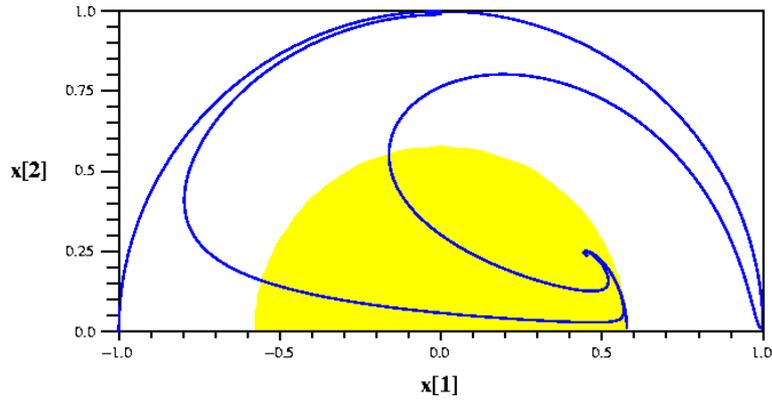}
\caption{Model with parameters $(\lambda,\gamma)=(\sqrt{2},4)$.}
\label{fig:phaseplot(sqrt{2},4)}
\end{center}
\end{figure}
\begin{figure}
\begin{center}
\includegraphics[width=10cm]{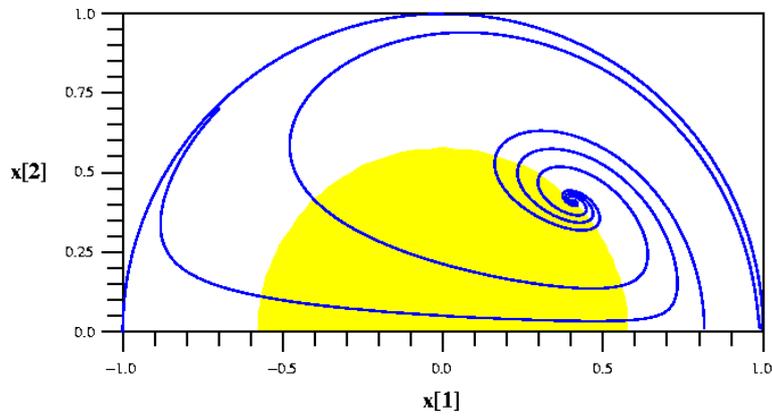}
\caption{Model with parameters $(\lambda,\gamma)=(2,4)$.}
\label{fig:phaseplot(2,4)}
\end{center}
\end{figure}

\newpage
\subsubsection{$J$ spiral focus, $G$ non-existing}
\label{STrec}

In this case, recurrent acceleration is most generically realized. At $\lambda=\sqrt{6},$ 
the fixed point $G$ merges with $B_1$. The dynamical evolution 
of the system is now totally determined by the saddle points $B_1$ and $B_2$, situated at the phase-space boundary, and the spiral focus $J$. (The condition (\ref{eq:spiral}) is trivially fulfilled in this part of parameter space.) Each trajectory winds around the attractor several times, undergoing subsequent stages of accelerated and decelerated expansion. As figures \ref{fig:phaseplot(3,6)} and \ref{fig:phaseplot(4,20)} show, this feature is almost independent of the position of the attractor with respect to the acceleration boundary. 

\begin{figure}[t]
\begin{center}
\includegraphics[width=10cm]{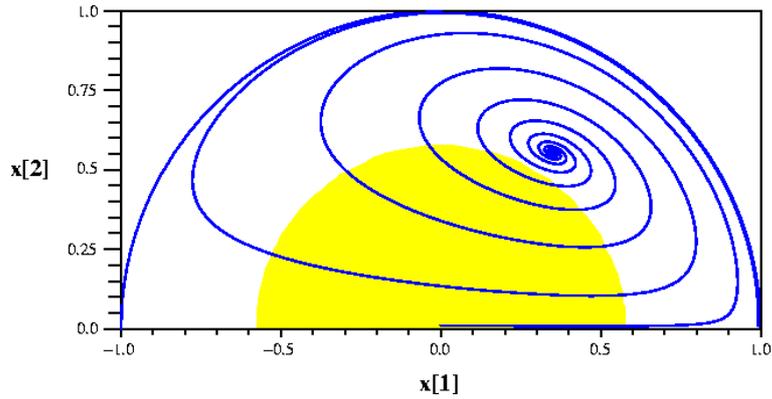}
\caption{Model with parameters $(\lambda,\gamma)=(3,4)$. Only two trajectories are shown.}
\label{fig:phaseplot(3,6)}
\end{center}
\end{figure}
\begin{figure}
\begin{center}
\includegraphics[width=10cm]{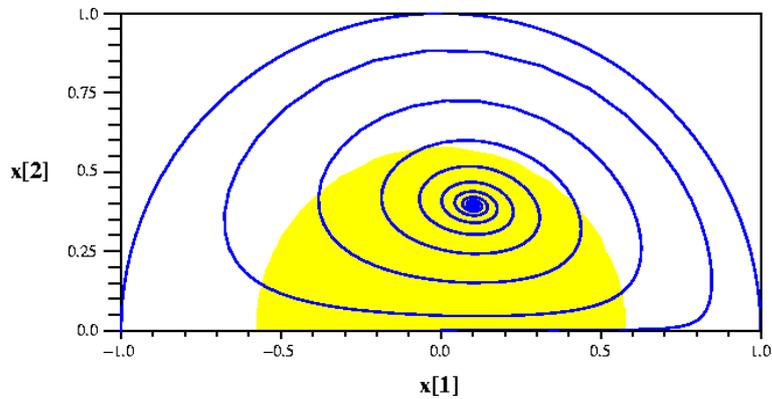}
\caption{Model with parameters $(\lambda,\gamma)=(4,20)$. Only a single trajectory is shown.}
\label{fig:phaseplot(4,20)}
\end{center}
\end{figure}

\newpage
\section{Axion-dilaton dynamics in presence of \\cosmological fluids: the case Q=0}
\label{Q0} 

We are now prepared to discuss stationary solutions of the full system (\ref{eqn:DSi})--(\ref{eqn:DSf}), but setting $Q=0$. We find the following set of critical points $X_s=(x_{1,s},x_{2,s},y_s,z_s)$:
\begin{align}
A &:\quad(0,0,0,0),\notag\\
B_1,B_2 &:\quad(\pm 1,0,0,0),\notag\\
C &:\quad(0,0,0,1),\notag\\
E &:\quad\left(\frac{2\sqrt{2}}{\sqrt{3}\lambda},0,
\frac{2}{\sqrt{3}\lambda},\sqrt{1-\frac{4}{\lambda^2}}\right),\notag\\
F &:\quad\left(\frac{\sqrt{3}}{\sqrt{2}\lambda},0,
\frac{\sqrt{3}}{\sqrt{2}\lambda},0\right),\notag\\
G &:\quad\left(\frac{\lambda}{\sqrt{6}},0,\sqrt{1-\frac{\lambda^2}{6}},0\right),\notag\\
J &:\quad\left(\frac{\sqrt{6}}{\gamma+\lambda},
\sqrt{\frac{\lambda(\gamma+\lambda)-6}{(\gamma+\lambda)^2}},\sqrt{\frac{\gamma}{\gamma+\lambda}}
,0\right).\notag
\end{align}

The density parameter of the AD system is given by
$$\Omega_{\textrm{AD}}=x_1^2+x_2^2+y.$$
Furthermore, we note that now, in the general case,
$$\frac{1}{\Omega_{\textrm{AD}}}(x_1^2+x_2^2-y^2)=w_{\textrm{AD}}\neq w_{\textrm{eff}}=x_1^2+x_2^2-y^2+\frac{1}{3}z^2.$$
The existence condition reads
$$x_1^2+x_2^2+y^2+z^2\leq 1,$$
with $x_1,x_2,y,z$ real. The eigenvalues of the Jacobi matrix 
determining stability of the different fixed points are given in the appendix. 
We display properties of the fixed points in table \ref{tab:Q=0}. 

\begin{table}[h]\begin{center}
\begin{tabular}{|c|c|c|c|c|}
\hline
\textbf{fixed point}&\textbf{existence}&\textbf{stability}
&\textbf{$\Omega_{\textrm{AD}}$}&\textbf{$w_{\textrm{eff}}$}\\
\hline\hline
$A$ & $\forall (\gamma,\lambda)$ & saddle point & 0 & 0\\[2pt]
\hline
$B_1,B_2$ & $\forall (\gamma,\lambda)$ & unstable & 1 & 1\\[2pt]
\hline
$C$ & $\forall (\gamma,\lambda)$ & unstable & 0 & $\frac{1}{3}$\\[2pt]
\hline
$E$ & $\lambda\ge 2$ & saddle point: & &\\
& & $\lambda>max\lbrace2,2\gamma\rbrace$ & $\frac{4}{\lambda^2}$ & $\frac{1}{3}$\\[2pt]
\hline
$F$ & $\lambda\geq\sqrt{3}$ & stable: & &\\
& & $\lambda>max\lbrace\sqrt{3},\gamma\rbrace$ &
$\frac{3}{\lambda^2}$ & $0$\\[2pt]
\hline
$G$ & $\lambda\leq \sqrt{6}$ & stable:  $\lambda<\sqrt{3}$ & &\\
 & & $\wedge~\lambda(\lambda+\gamma)<6$ & 1 & $-1+\frac{\lambda^2}{3}$\\[2pt]
\hline
$J$ & $\lambda(\lambda+\gamma) > 6$ & stable: & & \\
& $\wedge~\gamma\ge 0$ & $\gamma >\lambda$ & 1 & $\frac{\lambda-\gamma}{\lambda+\gamma}$\\[2pt]
\hline
\end{tabular}
\caption{Properties of the fixed points of system (\ref{eqn:DSi})--(\ref{eqn:DSf}), with $Q=0$.}
\label{tab:Q=0}
\end{center}\end{table}

We find a radiation dominated repeller $C$, a matter dominated saddle point $A$, 
and three different AD dominated regimes, $B_{1,2}, G,$ and $J,$ already present 
in the reduced system ($S_{\textrm{fluid}} = 0$). 
In addition, there are two different {\it scaling solutions}\footnote{
See \cite{Scaling} for a definition.}: 

$E$, where the energy density of the AD system scales like radiation, and $F,$ 
where it behaves like matter. These two fixed points are characterized by 
$\Omega_{\textrm{AD}}<1$. 

The subset of fixed points exhibiting $x_{2}=0$ is identical to the set of fixed 
points characterizing single-field models with exponential potential (see \cite{DEreview}, and references therein). 
These fixed points correspond to trivial solutions of the axion equation of motion (\ref{eoma}) and
our analysis shows that such configurations are indeed stable in a wide range of parameter space.

The existence of stationary solutions with $x_{2}\neq 0$ is related to the sign of the friction
term in the axion equation of motion, {\it i.e.} the quantity $3H-\gamma\dot{\Phi}$.
As long as $3H~-~\gamma\dot{\Phi}~>~0,$ the axion evolves toward a configuration where
$\dot{\sigma}=0$ and therefore $x_{2}=0$. On the other hand, stability of the fixed point $J$  
with $x_{2}\neq 0$ implies $x_{1}|_{J}>\gamma^{-1}\sqrt{3/2}$, which is equivalent to 
$3H-\gamma\dot{\Phi}<0$.

\begin{figure}[hpt]
\begin{center}
\includegraphics[width=10cm]{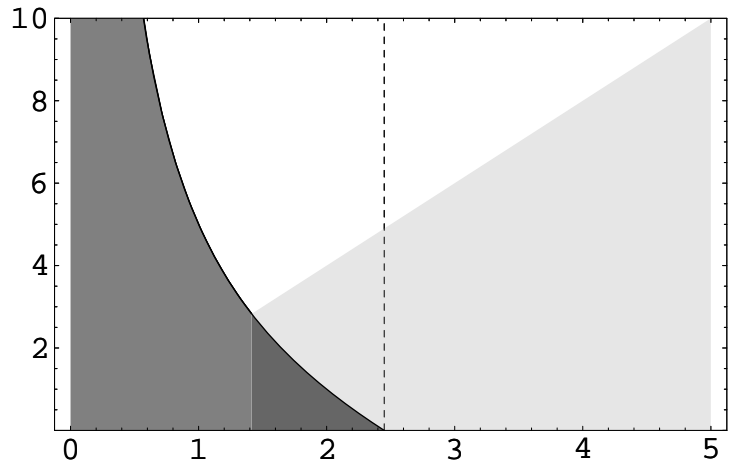}
\caption{Parameter space of the Sonner-Townsend family of models. 
Horizontal axis: $\lambda$, vertical axis: $\gamma$. In the gray region the fixed 
point $G$ is stable. The dashed line is the existence boundary of $G$. 
In the shaded region acceleration is impossible at the attractor.}
\label{STparam}
\end{center}
\end{figure}

\begin{figure}[hp]
\begin{center}
\includegraphics[width=10cm]{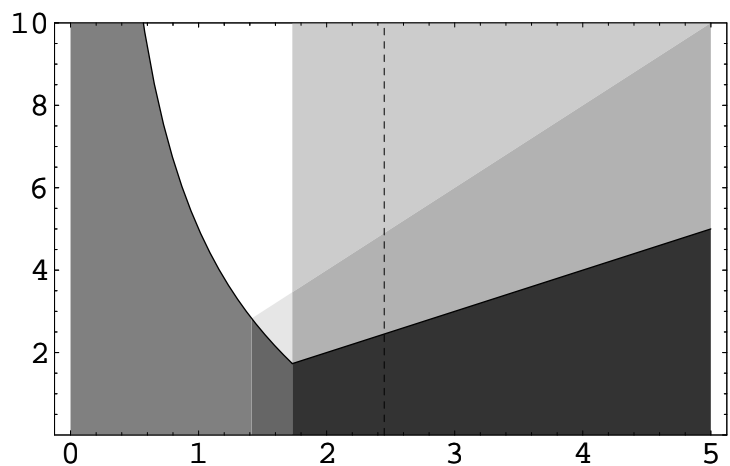}
\caption{As above, now the general case. The existence of the new fixed point $F$ 
is indicated for $\lambda\geq \sqrt{3},$ in the dark gray region $F$ is the attractor.}
\label{ADparam}
\end{center}
\end{figure}

In figures \ref{STparam} and \ref{ADparam} we show the different domains of stability in 
parameter space of both theories, with and without a cosmological fluid background. 
Most relevant is the appearance of the new fixed point $F$, which is either a stable 
focus or a saddle point in a significant range of parameter space.


\subsection{Recurrent acceleration in presence of background fluids?}

In this subsection we investigate to which extent a perfect fluid background 
affects the possibility of recurrent acceleration.

Following our discussion presented in section~\ref{nofluid}, we assume $J$ to be a spiral focus, located close enough to the acceleration boundary in phase-space. In other words, we restrict ourselves to a subclass of models, corresponding to the neighborhood of the line $\gamma = 2\,\lambda$ in parameter space. We have to discuss three different cases in turn.
\begin{itemize}
\item $\lambda <\sqrt{2}$: $G$ saddle point, accelerated expansion at $G$;
\item $\sqrt{2} <\lambda\leq\sqrt{3}$: $G$ saddle point, deceleration at $G$;
\item $\lambda >\sqrt{3}$: $F$ saddle point.
\end{itemize}
We will focus on the behavior of the trajectory connecting the saddle point ($F$ or $G$)
with the attractor $J$. Generalizing our previous definition, hereafter we will refer to such a
trajectory as connecting trajectory.

Starting with the first case, we note that
$w_{\textrm{eff}}(G)> w_{\textrm{eff}}(J)$.
This follows from the existence condition of $J$, which can be rewritten as
$\lambda^2 / 3> 2\lambda (\lambda+\gamma)^{-1}.$
The connecting trajectory is entirely contained not only within the domain of accelerated expansion, but also in the section of the phase space boundary defined by $\Omega_{\textrm{AD}}=1.$ 
Therefore, in this case we find no qualitative difference with respect to section~\ref{nofluid}. (See figure \ref{fig:phaseplot(sqrt{2},4)} for comparison.)

In the second case, the limiting trajectory itself crosses the acceleration boundary while spiraling onto the attractor. (See figure \ref{fig:phaseplot(2,4)} for comparison.) In particular, any trajectory which enters the acceleration domain {\it before} approaching the saddle point $G$ will experience at least two distinct stages of accelerated expansion. The first stage corresponds to the well-known {\it freezing} regime \cite{DEreview} of single-field models of dark energy: the dilaton field velocity remains close to zero due 
to the Hubble friction term dominating the equation of motion. The second stage is reached, when the trajectory re-enters the domain of acceleration in approaching the connecting trajectory and the regime of the late time attractor. If $J$ is located inside the domain, accelerated expansion will continue forever.

In the third case the saddle point $F$ is dynamically relevant.
Moreover, at $F$ $w_{\textrm{eff}}~=~w_{\textrm{AD}}~=~0$ and
the expansion is either dominated by matter or by the AD system scaling like matter, 
with $\Omega_{\textrm{AD}}<1.$ Due to the finite contribution of matter to the total energy density of the universe, we have now -- in contrast to the previous cases~--
$w_{\textrm{eff}}>w_{\textrm{AD}}$
along the connecting trajectory. Thus, even if the connecting trajectory 
oscillates around $w_{\textrm{AD}}(J)\approx -1/3$ 
before reaching the spiral focus, recurrent acceleration is not implied. In particular, in any model with 
$w_{\textrm{eff}}(J)\geq -1/3$ it will never enter the domain of acceleration at all. 
However, we cannot exclude the possibility of $w_{\textrm{eff}}$ 
crossing the acceleration boundary more than once, if at the attractor 
$w_{\textrm{eff}}(J) < -1/3$. 

\subsection{Numerical examples}

The conclusions we drew in the previous subsection can be circumvented by resorting to a very special choice of initial conditions. For instance, the scenario of \cite{Sonner} 
can be recovered by setting 
$$z_{\textrm{in}} = 0,\quad (x_1^2+x_2^2+y^2)_{\textrm{in}}=1.$$ 
However, since we are ultimately interested in models which are able to reproduce qualitatively the standard evolution of the universe, as it can be reconstructed from cosmological observations \cite{DEreview}, we will hereafter only consider trajectories which are (at least marginally) consistent with the concordance $\Lambda$CDM cosmology. Furthermore, we will henceforth identify $\Omega_{\textrm{AD}}$ with $\Omega_{\textrm{DE}}$
and $w_{\textrm{AD}}$ with $w_{\textrm{DE}}$, where the subscript DE refers to dark energy.

It is a well-known, serious problem of dynamical dark energy models that their late time evolution typically still depends on initial conditions: In single-field models, for instance, the energy scale of the potential, {\it i.e.} in our notation $y_{\textrm{in}}$, has to be fine tuned to satisfy $\Omega_{\textrm{DE,today}}\approx 0.75.$ 
In the present case one has to deal with an additional sensitivity on  $x_{2,\textrm{in}}$, which will be explained below.

We will discuss these issues on the basis of two numerical examples, which are both characterized by an attractor solution given by fixed point $J,$ preceded by a saddle point, which is $G$ in the first case, and in the second one $F$.  
We have specified the respective trajectory by imposing initial conditions for the AD system at a temperature of $\mathcal{O}(1)$ MeV 
when $(\Omega_{\textrm{rad}}/\Omega_{\textrm{mat}})_{\textrm{in}} \sim 10^{6}$, thereby ensuring the validity of our classical description.
We assume the scalar fields to have already reached the freezing regime, relying on the fact that scalar field kinetic energy scales as $a^{-6}$. This leaves the possibility of a stage of {\it kination} \cite{Dilrun} during a preceding epoch of higher temperature. 

\begin{figure}
\begin{center}
\includegraphics[width=12cm]{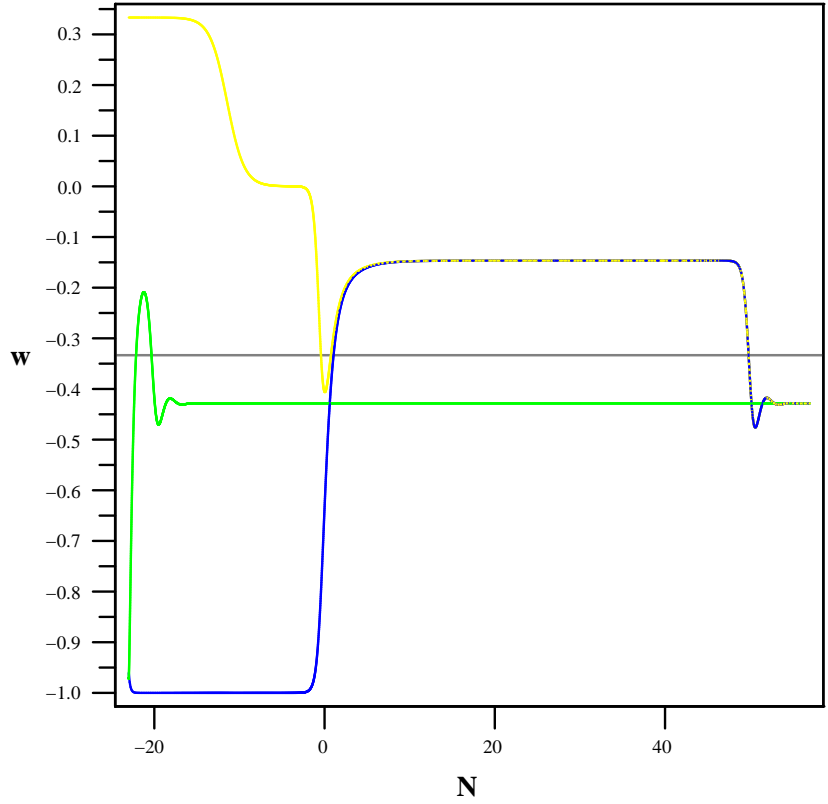}
\caption{Model with parameters $(\lambda,\gamma)=(\frac{8}{5},4)$. Evolution of $w_{\textrm{DE}}$ (blue, dark curve), $w_{eff}$ (yellow, light curve), with $N:= -\ln(1+z),\,N=0$ referring to the present. The trajectory is specified by initial conditions $(x_1,x_2,y)_{\textrm{in}}=(10^{-18}, 10^{-18}, 10^{-17})$. For comparison, the green curve shows the evolution along a trajectory with $\Omega_{\textrm{mat}}~=~\Omega_{\textrm{rad}}~=~0$, as in the Sonner Townsend case, but the same ratio between initial kinetic and potential energy of the scalar fields.}
\label{fig:EvolutionJ(G)}
\end{center}
\end{figure}

\begin{figure}
\begin{center}
\includegraphics[width=12cm]{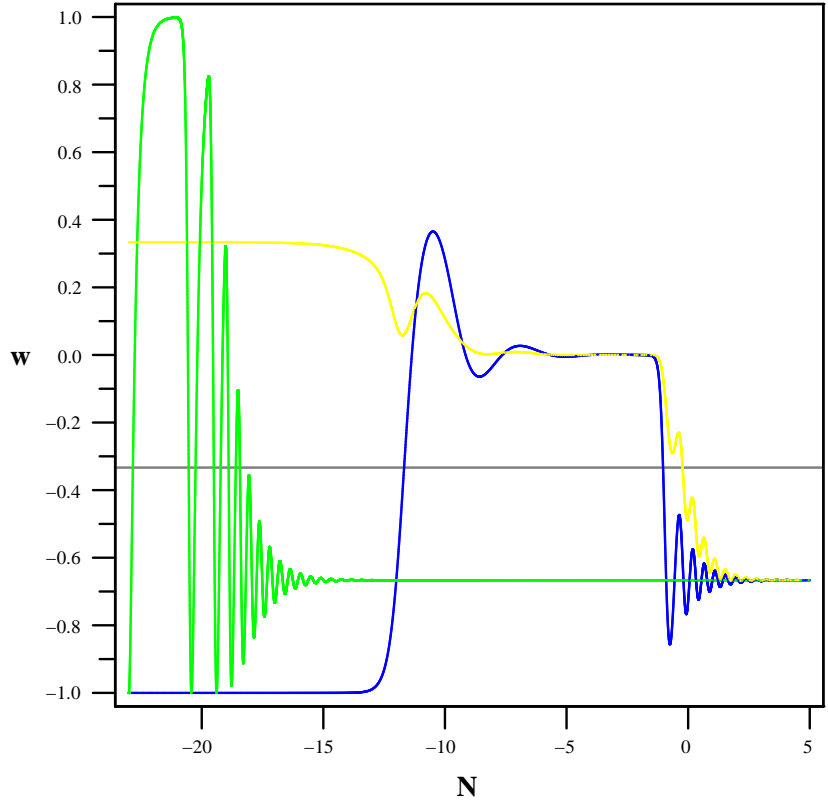}
\caption{Model with parameters $(\lambda,\gamma)=(4,20)$. Evolution of $w_{\textrm{DE}}$ (blue, dark curve), $w_{\textrm{eff}}$ (yellow, light curve), with $N:= -\ln(1+z),\,N=0$ referring to the present. The trajectory is specified by initial conditions $(x_1,x_2,y)_{\textrm{in}}=(5\times 10^{-28},5\times 10^{-28}, 5\times 10^{-10})$. For comparison, the green curve shows the evolution along a trajectory with $\Omega_{\textrm{mat}}= \Omega_{\textrm{rad}} = 0$, as in the Sonner Townsend case, but the same ratio between initial kinetic and potential energy in the scalar field sector.}
\label{fig:EvolutionJ(F)}
\end{center}
\end{figure}

As it can be seen in figures ~\ref{fig:EvolutionJ(G)} and \ref{fig:EvolutionJ(F)},
we find, in both cases, three successive evolutionary stages of the AD system: The first one, the freezing regime, 
is associated with the radiation dominated epoch (RDE) if $F$ is the saddle point, or continues during the matter dominated epoch (MDE), if $G$ is the saddle point.  Thereafter the system enters the regime of the saddle point, which in both cases lasts for a significant number of $e$-foldings of expansion.
Finally, there is the late time attractor regime. Remarkably, our present situation 
corresponds to the transition between stage one and two in one case (saddle point $G$),
and two and three in the other (saddle point $F$). 
This is related to the significant discrepancy in $y_{\textrm{in}}$ 
in our two examples (see the figure captions). Changes in $y_{\textrm{in}}$ affect in particular the termination of the freezing regime, while the choice of $x_{2,\textrm{in}}$ determines the duration of the saddle point regime. 

The value of $x_{2}$ decreases monotonically along a given trajectory as long as $x_1 <\sqrt{3/2}\,\gamma^{-1}$ 
and turns to increase when $x_1 > \sqrt{3/2}\,\gamma^{-1}$ (which is already true at the saddle point $F$). 
The saddle point regime ceases, once $x_2$ has increased sufficiently to perturb $x_1$ away from its fixed point value: 
A non-zero axion kinetic term contributes to the effective potential in the dilaton equation of motion, Eq. (\ref{eomd}).

On the contrary, the initial condition $x_{1, \textrm{in}}$ influences just the early stage of
the dynamical evolution: A set of trajectories differing only in $x_{1,\textrm{in}}$ first converge toward the saddle point before they start to approach the attractor, thereby washing out any dependence of the late time evolution on $x_{1,\textrm{in}}$. The corresponding evolutionary path in phase-space is therefore completely determined by the 
connecting trajectory, {\it provided} we can safely assume $x_{2,\textrm{in}}$ to be sufficiently small.

To demonstrate the impact of a cosmological background fluid on the AD dynamics, in particular 
concerning the phenomenon of recurrent acceleration present in the case $S_{\textrm{fluid}}=0,$ we have also plotted -- for comparison -- the evolution of the system's equation of state in absence of the fluid background
(observe the green lines in the figures). The admixture of a perfect 
fluid component to the initial composition has two effects on the evolution of the scalar fields:
First, the reduction of $\Omega_{\textrm{DE},\textrm{in}}$ 
by a huge factor $\mathcal{O} (10^{-10})$ allows for the existence of a freezing regime during RDE and/or MDE. Secondly (see figure \ref{fig:EvolutionJ(F)}), due to the existence of the scaling saddle point $F$, the first few, large amplitude oscillations of the equation of state are suppressed and partly replaced by oscillations around the saddle point value, leaving only rapid, small amplitude oscillations around the attractor value. Needless to say, this kind of oscillations in the DE and effective equation of state are -- at low redshift -- already disfavored by observational data \cite{DEreview, Alam}.

We have to conclude that the characteristic feature of recurrent acceleration, 
as illustrated in figure \ref{fig:phaseplot(4,20)}, disappears if we allow 
for a perfect fluid contribution dominating the earlier stages of evolution. 
In particular, the existence of a scaling solution preceding the 
spiral focus regime reduces number, amplitude and period of possible oscillations in $w_{\textrm{eff}}$ crossing the acceleration boundary. 

\newpage
\section{Axion-dilaton dynamics in presence of \\cosmological fluids: the case $Q > 0$}
\label{Q}

The set of equations (\ref{eqn:DSi})--(\ref{eqn:DSf}), with $Q$ allowed to be non-zero, defines a three-parameter family of models, each characterized by a set of fixed points in a four-dimensional compact phase-space. These fixed points are:
\begin{align}
A : &\quad\left(\sqrt{\frac{2}{3}}Q,0,0,0\right),\notag\\
B_1,B_2: &\quad(\pm 1,0,0,0),\notag\\
C: &\quad(0,0,0,1),\notag\\
D: &\quad\left(\frac{1}{\sqrt{6}Q},0,0,\sqrt{1-\frac{1}{2Q^2}}\right),\notag\\
E: &\quad\left(\frac{2\sqrt{2}}{\sqrt{3}\lambda},0,
\frac{2}{\sqrt{3}\lambda},\sqrt{1-\frac{4}{\lambda^2}}\right),\notag\\
F: &\quad\left(\frac{\sqrt{\frac{3}{2}}}{\lambda-Q},0,
\sqrt{\frac{2Q(Q-\lambda)+3}{2(\lambda-Q)^2}},0\right),\notag\\
G: &\quad\left(\frac{\lambda}{\sqrt{6}},0,\sqrt{1-\frac{\lambda^2}{6}},0\right),\notag\\
H: &\quad\left(\frac{\sqrt{\frac{3}{2}}}{\gamma+Q},
\sqrt{\frac{2Q(\gamma+Q)-3}{2(\gamma+Q)^2}}
,0,0\right),\notag\\
J: &\quad\left(\frac{\sqrt{6}}{\gamma+\lambda},
\sqrt{\frac{\lambda(\gamma+\lambda)-6}{(\gamma+\lambda)^2}},
\sqrt{\frac{\gamma}{\gamma+\lambda}},0\right).\notag
\end{align}

With respect to the case $Q=0$, we find two additional stationary points, namely $D$ and $H$. 
The fixed point $D$ is associated to RDE.  
Properties of the fixed points are displayed in table \ref{tab:Q}.

\begin{table}[h]\begin{center}
\begin{tabular}{|c|c|c|c|c|}
\hline
\textbf{f.p.}&\textbf{existence}&\textbf{stability}
&\textbf{$\Omega_{DE}$}&\textbf{$w_{eff}$}\\
\hline\hline
$A$ & $Q\le\sqrt{\frac{3}{2}}$ & stable: & &\\
 & & $Q^2< min\lbrace \frac{1}{2},\frac{3}{2}-\gamma Q, \lambda Q -\frac{3}{2}\rbrace$ & $\frac{2}{3}Q^2$ & $\frac{2}{3}Q^2$\\[2pt]
\hline
$B_1$ &  & saddle point: & & \\
& $\forall (\gamma,\lambda, Q)$ & $ \lambda>\sqrt{6}~\wedge~Q>\sqrt{\frac{3}{2}}~\wedge~\gamma<0$ & 1 & 1\\
$B_2$ & & unstable & &\\[2pt]
\hline
$C$ & $\forall (\gamma,\lambda, Q)$ & unstable & 0 & $\frac{1}{3}$\\[2pt]
\hline
$D$ & $Q\ge\frac{1}{\sqrt{2}}$ & stable: $\lambda>4Q>2\gamma$ & $\frac{1}{6Q^2}$ & $\frac{1}{3}$\\[2pt]
\hline
$E$ & $\lambda\ge 2$ & stable: $2\gamma<\lambda<4Q$ & $\frac{4}{\lambda^2}$ & $\frac{1}{3}$\\[2pt]
\hline
$F$ & $\frac{3}{2Q}+Q\geq\lambda\geq\frac{Q+\sqrt{Q^2+12}}{2}$ & stable: $Q<\frac{1}{\sqrt{2}}$ & &\\
 & $\wedge\,Q\leq\sqrt{\frac{3}{2}}$ & $\wedge ~ \lambda> max\lbrace 4Q, 2Q+\gamma\rbrace$ & $\frac{3+Q^2-Q\lambda}{(Q-\lambda)^2}$ & $\frac{Q}{\lambda-Q}$\\[2pt]
\hline
$G$ & $\lambda\le \sqrt{6}$ & stable: & &\\
 & & $\lambda^2<min\lbrace 4,3+Q\lambda,6-\gamma\lambda \rbrace$ & 1 & $-1+\frac{\lambda^2}{3}$\\[2pt]
\hline
$H$ & $\gamma\ge max\lbrace 0,\frac{3}{2Q}-Q\rbrace$ & stable: $\lambda>\gamma+2Q$ & &\\
 & & $\wedge~\gamma> 2Q$ & $\frac{Q}{\gamma+Q}$ & $\frac{Q}{\gamma+Q}$\\[2pt]
\hline
$J$ & $\lambda(\lambda+\gamma) > 6$ & stable: & &\\
& $\wedge~\gamma\ge 0$ & $\lambda< min\lbrace 2\gamma, \gamma+2Q\rbrace$ & 1 & $\frac{\lambda-\gamma}{\lambda+\gamma}$\\[2pt]
\hline
\end{tabular}
\caption{Properties of the fixed points in the case $Q > 0.$}
\label{tab:Q}
\end{center}\end{table}

\begin{figure}[hpt]
\begin{center}
\includegraphics[width=10cm]{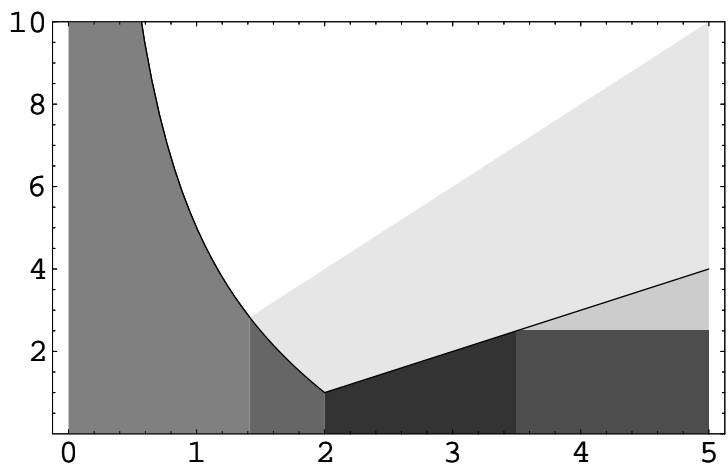}
\caption{Parameter-space in the case $Q=\frac{1}{2}$. Horizontal axis: $\lambda$, vertical axis: $\gamma$. Regions of stability of the various fixed points are indicated. The full line is the stability boundary of $J$. $F$ can only be stable if $2<\lambda<7/2.$ For $\lambda>7/2$ we find new attractors $A$ (if $\gamma<5/2$) and $H$. In the shaded region acceleration is impossible at the attractor.}
\label{fig:paramQ}
\end{center}
\end{figure}

\begin{figure}[hp]
\begin{center}
\includegraphics[width=10cm]{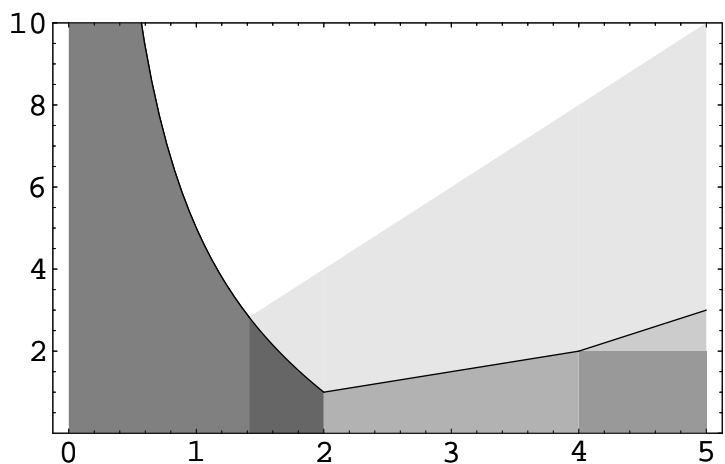}
\caption{As above, now the case $Q=1.$ Attractors $F$ and $A$ are replaced by $E$ (if $2<\lambda<4$) and $D$, both corresponding to radiation era.}
\label{fig:paramQ1}
\end{center}
\end{figure}

In figures \ref{fig:paramQ} and \ref{fig:paramQ1} 
we show two sections of parameter space, with $Q=1/2$ and $Q=1$ respectively, 
to cover the different possibilities of stable fixed points. 
Due to the positivity of $Q$, the AD energy density gets enhanced 
at expense of the matter sector. If $Q$ is sufficiently large, not 
a single fixed point remains with $w_{\textrm{eff}}$ equal or at least close 
to zero, indicating suppression of MDE. However, we have to note that 
such large values of $Q$ are unphysical because of the existing bounds
on a universal metric coupling between matter and gravity \cite{Damour, STconstraints}. 

Moreover, let us emphasize that increasing $Q$ does not re-establish recurrent 
acceleration. As in the case $Q=0$, generic trajectories which 
converge toward the spiral focus $J$ approaching the connecting trajectory, 
always spend a certain number of $e$-foldings 
close to a saddle point, where either radiation or matter dominate. 
Depending on the parameter values, the relevant saddle point is either $F$, $A$, $H$, $D$ or $E$. 


\newpage
\section{Discussion and conclusions} 
\label{Conclusions}

We have investigated a family of flat FRW cosmological models in $D=4$, focusing on the dynamics of a scalar (dilaton) and a pseudo-scalar (axion) partner of the metric field in presence of cosmological background fluids. Neglecting any specific interaction term of the axion and the dilaton, which are present in particle physics or string motivated models, we have considered the -- simplified -- scenario of universal metric coupling. Motivated by recent results of \cite{Sonner}, where recurrent acceleration was verified in a large class of AD models {\it in absence} of matter or radiation, we have analyzed the possibility to combine the phenomenon of recurrent acceleration with a cosmological background evolution in concordance with $\Lambda$CDM cosmology. In particular, we have considered the viability of the AD system as a candidate of dynamical dark energy.

Using a different choice of dynamical variables, well-suited to the more general case with background fluids, we were able to reproduce the findings of \cite{Sonner} (corresponding to $S_{\textrm{fluid}} = 0$) . The new feature of models with $S_{\textrm{fluid}} \neq 0$ is the existence of a stationary cosmological scaling solution within a wide range of parameter space. As long as $Q$ is (close to) zero, the relevant fixed point is $F$, corresponding to matter dominance. In particular, if being a saddle point, $F$ influences the evolution along generic trajectories in such a way that the phenomenon of recurrent acceleration, as observed in \cite{Sonner}, is reduced to small amplitude oscillations of the equation of state at low redshift.

We can certainly realize a model (by specifying parameters), which is able to reproduce the present stage of cosmic evolution as being a {\it transient} phenomenon. This is already possible in the single-field case, corresponding to the constant axion scenario within our dynamical system: The dilaton remains frozen, due to Hubble friction dominance, right up to the present, and later-on evolves toward an attractor solution exhibiting $w_{\textrm{eff}}>-1/3.$ On the other hand, if the axion dynamics is non-trivial, it is also possible to get a second accelerating stage in the future, which will then be ever-lasting.

Though we have to conclude that recurrent acceleration as described in \cite{Sonner} is not relevant to dark energy model building, we have discovered a different, interesting possibility instead, which we may call the $F\rightarrow J$ scenario. Does it provide a viable model of dynamical dark energy?

The co-existence of a matter-dominated scaling solution and a dark energy dominated accelerating solution is generally considered to be a very attractive feature of dynamical DE models \cite{DEreview, Scaling}. In the single-field case \cite{CLW}, the existence of $F$ is excluded by the stability condition of the fixed point $G$, which is the only available candidate to achieve late time acceleration. However, in presence of an axion field, the scaling regime of $F$ can be succeeded by a stage of accelerated expansion, represented by the fixed point $J$. During radiation and matter dominated epochs, the evolution of the AD sector is determined by the saddle point, while our present situation corresponds to the transition toward the late time attractor solution. Unfortunately, the onset of this transition is subject to a certain fine tuning of initial conditions.

One single fine tuning of the scalar potential energy scale is always mandatory in dynamical models of DE, since it corresponds to setting the cosmic clock. The requirement of naturalness \cite{Wetterich} strongly disfavors models which need a fine tuning of the same order as in the cosmological constant case. In this respect, two-field models incorporating the $F\rightarrow J$ scenario are certainly promising: In our numerical example, the potential scale is initially set to $\mathcal{O} (10^{15})\times \Omega_{\Lambda}|_{1\textrm{MeV}}$. Even larger values are possible, but have to be compensated by reducing $x_{2,\textrm{in}},$ in order to keep the cosmic clock tuned. 
We need to emphasize, however, that extremely small values of the axion field velocity are by no means unphysical. In fact, the most natural solution is a constant axion, as long as the Hubble rate is large enough to keep the friction term positive. 

To conclude, let us stress an intriguing feature of the class of models under consideration. Single-field potentials are typically required to be extremely flat in order to satisfy observational constraints. However, in the $F\rightarrow J$ scenario this is not true: there is no upper bound on the parameter $\lambda$ determining the potential slope. This aspect opens up new possibilities for dark energy model
building which we believe to be promising enough to motivate further investigation. 


\section*{Acknowledgments}
We would like to thank Wilfried Buchm\"uller for enlightening comments
and suggestions during the development of this work.
R. Catena acknowledges a Research Grant funded by the VIPAC Institute.

\newpage
\appendix
\section{Dynamical systems terminology}

We consider a system of $n$ first order ordinary differential equations (ODE),
\begin{equation}
\dot{x}_i=f_i(x_1,...,x_n),
\end{equation}
which is called {\itshape autonomous} if none of the $n$ functions $f_i$ explicitly depends on time. A solution of the system is given in terms of a trajectory in phase space,
$$t\,\longmapsto\,X(t):=(x_1(t),...,x_n(t)),$$
determined by choice of initial conditions $X(t_{init})$.

A point $X_s:=(x_{1,s},...,x_{n,s})$ is said to be a {\itshape critical, stationary} or {\itshape fixed point} if
$$f_i(X_s)=0 \qquad \forall \, i\,\leq\, n,$$
and an {\itshape attractor} if there exists a neighborhood of the fixed point such that every trajectory entering this neighborhood satisfies the following condition:
$$\lim_{t\rightarrow\infty} X(t)= X_s.$$

Now consider small perturbations around the critical point,
$$x_i=x_{i,s}+\delta x_i.$$
Linearizing the evolution equations we obtain a system of first order ODE {\itshape linear} in the perturbations,
\begin{equation}
\frac{d}{dt}\delta x_i=\sum_j M_{ij}\delta x_j,
\end{equation}
where
$$M_{ij}:=\left.\frac{\partial f_i(X)}{\partial x_j}\right|_{X=X_s}.$$
The general solution of this system is given by
$$\delta x_i=\sum_{k=1}^{n} C_{ik} e^{\mu_k t}, $$
where $C_{ik}$ are integration constants and $\mu_k$ the eigenvalues of the Jacobi or {\itshape stability matrix} $M$, which we have assumed to be distinct for simplicity. Obviously the perturbation will decay if each $\mu_k$ has negative real part.

The critical points of a dynamical system can be classified in terms of the eigenvalues of the corresponding stability matrix. An attractor is characterized by the requirement
$$Re[\mu_k]<0 \qquad \forall\, k\,\leq\, n,$$ 
and called {\itshape spiral focus} if at least one pair of eigenvalues is complex and {\itshape stable node} else. Furthermore we will use the terminus {\itshape saddle point} if and only if $M$ has one eigenvalue with positive real part. In any other case we call the fixed point {\itshape unstable}.

\section{Jacobi matrix eigenvalues}

In the following we list the Jacobi matrix eigenvalues at the different fixed points
of the models with background fluids.

\subsection{Case $Q=0$}

$$A:\quad -\frac{3}{2},\,-\frac{3}{2},~\frac{3}{2},\,-\frac{1}{2} ,$$

$$B_1:\quad 3,~1,~\sqrt{\frac{3}{2}}\gamma,~3-\sqrt{\frac{3}{2}}\lambda,$$

$$B_2:\quad 3,~1,\,- \sqrt{\frac{3}{2}}\gamma,~3+\sqrt{\frac{3}{2}}\lambda,$$

$$C:\quad 2,\,-1,\,-1,~1,$$

$$E:\quad 1,\,-1+\frac{2\gamma}{\lambda},
~\frac{1}{2}\left(-1\pm\frac{\sqrt{64\lambda^2-15\lambda^4}}{\lambda^2}\right),$$

$$F:\quad -\frac{1}{2},~\frac{3(\gamma-\lambda)}{2\lambda},
~\frac{3}{4}\left(-1\pm\frac{\sqrt{24\lambda^2-7\lambda^4}}{\lambda^2}\right),$$

$$G:\quad\frac{1}{2}(\lambda^2 -6),~ \frac{1}{2}(\lambda(\lambda+\gamma)-6),
~\lambda^2-3,~\frac{1}{2}(\lambda^2 -4),$$

$$J:\quad 3\left(1-\frac{2\gamma}{\gamma+\lambda}\right),~1-\frac{3\gamma}{\gamma+\lambda},$$
$$\frac{3}{2(\gamma+\lambda)}\left(-\gamma \pm
\sqrt{\gamma^2+8\gamma(\gamma+\lambda)-\frac{4}{3}\gamma\lambda(\gamma+\lambda)^2}\right).$$

\newpage
\subsection{Case $Q\neq 0$}
$$A:\quad -\frac{1}{2}+Q^2,\,-\frac{3}{2}+Q^2,\,-\frac{3}{2}+Q(Q+\gamma),
~\frac{3}{2}+Q(Q-\lambda),$$

$$B_1:\quad 1,~3-\sqrt{6}Q,~\sqrt{\frac{3}{2}}\gamma,~3-\sqrt{\frac{3}{2}}\lambda,$$

$$B_2:\quad 1,~3+\sqrt{6}Q,\,- \sqrt{\frac{3}{2}}\gamma,~3+\sqrt{\frac{3}{2}}\lambda,$$

$$C:\quad 2,\,-1,\,-1,~1,$$

$$D:\quad -1+\frac{\gamma}{2Q},~2-\frac{\lambda}{2Q},~\frac{1}{2}
\left(-1\pm\frac{\sqrt{2Q^2-3Q^4}}{Q^2}\right),$$

$$E:\quad 1-\frac{4Q}{\lambda},\,-1+\frac{2\gamma}{\lambda},
~\frac{1}{2}\left(-1\pm\frac{\sqrt{64\lambda^2-15\lambda^4}}{\lambda^2}\right),$$

$$F:\quad -\frac{\lambda-4Q}{2(\lambda-Q)},~\frac{3}{2}\left(-1+\frac{\gamma+Q}{\lambda-Q}\right),$$
$$\frac{3}{4(\lambda-Q)^2}\left(-(\lambda-2Q)(\lambda-Q)\pm
\sqrt{(\lambda-Q)^2[24-7\lambda^2-12\lambda Q+20Q^2+\frac{16}{3}\lambda Q(\lambda-Q)^2]}
\right),$$

$$G:\quad\frac{1}{2}(\lambda^2 -6),~ \frac{1}{2}(\lambda(\lambda+\gamma)-6),
~-3+\lambda(\lambda-Q),~\frac{1}{2}(\lambda^2 -4),$$

$$H:\quad 1-\frac{3\gamma}{2(\gamma+Q)},~\frac{3}{2}(1-\frac{\lambda-Q}{\gamma+Q}),$$
$$\frac{1}{4(\gamma+Q)}\left(-3\gamma\pm
\sqrt{81\gamma^2-24\gamma Q[2(\gamma+Q)^2-3]}\right),$$

$$J:\quad 3\left(1-\frac{2(\gamma+Q)}{\gamma+\lambda}\right),~1-\frac{3\gamma}{\gamma+\lambda}.$$
$$\frac{3}{2(\gamma+\lambda)}\left(-\gamma \pm
\sqrt{\gamma^2+8\gamma(\gamma+\lambda)-\frac{4}{3}\gamma\lambda(\gamma+\lambda)^2}\right)$$


\begin{thebibliography}{99}

\bibitem{moduliproblem}
  G.~D.~Coughlan, W.~Fischler, E.~W.~Kolb, S.~Raby, and G.~G.~Ross,
  Phys.\ Lett.\ B {\bf 131}, 59 (1983);
  B.~de Carlos, J.~A.~Casas, F.~Quevedo, and E.~Roulet,
  Phys.\ Lett.\ B {\bf 318}, 447 (1993);
  S.~B.~Giddings, S.~Kachru, and J.~Polchinski,
  Phys.\ Rev.\  D {\bf 66}, 106006 (2002); 
  S.~Kachru, R.~Kallosh, A.~Linde, and S.P.~Trivedi,
  Phys.~Rev. {\bf D68}, 046005 (2003);
  M.~Endo, K.~Hamaguchi, and F.~Takahashi,
  Phys.\ Rev.\ Lett.\  {\bf 96}, 211301 (2006);
  J.~P.~Conlon and F.~Quevedo,
  arXiv:0705.3460[hep-ph].
\bibitem{DEreview}
  E.~J.~Copeland, M.~Sami, and S.~Tsujikawa,
  Int. J. Mod. Phys. D {\bf 15}, 1753 (2006).
\bibitem{ST}
  P.~Jordan, {\em Schwerkaft und Weltall} (Vieweg, Braunschweig,
  1955); M. Fierz, Helv. Phys. Acta {\bf 29}, 128 (1956);
  C.~Brans and R.~H.~Dicke,
  Phys.\ Rev.\ {\bf 124}, 925 (1961);
  T.~Damour and K.~Nordtvedt,
  Phys.\ Rev.\ Lett.\  {\bf 70}, 2217 (1993);
  R.~Catena, M.~Pietroni and L.~Scarabello,
  arXiv:astro-ph/0604492.
\bibitem{ST1}
  R.~Catena, N.~Fornengo, A.~Masiero, M.~Pietroni and F.~Rosati,
  Phys.\ Rev.\  D {\bf 70}, 063519 (2004),
  R.~Catena, M.~Pietroni and L.~Scarabello,
  Phys.\ Rev.\  D {\bf 70}, 103526 (2004),
  M.~Schelke, R.~Catena, N.~Fornengo, A.~Masiero and M.~Pietroni,
  Phys.\ Rev.\  D {\bf 74}, 083505 (2006).
\bibitem{Damour}  
  T.~Damour,
  arXiv:gr-qc/9606079,
  lectures given at Les Houches Summer School on Gravitation and Quantizations.
\bibitem{STDE} 
  N.~Bartolo and M.~Pietroni,
  Phys.\ Rev.\ D {\bf 61} (2000) 023518;
  G.~Esposito-Farese and D.~Polarski,
  Phys.\ Rev.\ D {\bf 63}, 063504 (2001);
  R.~Gannouji, D.~Polarski, A.~Ranquet and A.~A.~Starobinsky,
  JCAP {\bf 0609}, 016 (2006).
\bibitem{inflation}
  J.~Garcia-Bellido and D.~Wands,
  Phys.\ Rev.\  D {\bf 52}, 6739 (1995).
\bibitem{Witten}
  M.B. Green, J.H. Schwarz, and E. Witten. 
  {\it Superstring Theory} (Cambridge University Press, Cambridge, 1987).
\bibitem{blackhole}  
  R.~Kallosh, A.~D.~Linde, T.~Ortin, A.~W.~Peet, and A.~Van Proeyen,
  Phys.\ Rev.\  D {\bf 46}, 5278 (1992);
  T.~Ortin,
  Phys.\ Rev.\  D {\bf 47}, 3136 (1993);
  R.~Kallosh and T.~Ortin,
  Phys.\ Rev.\  D {\bf 48}, 742 (1993).
\bibitem{domainswall}
  J.~Sonner and P.~K.~Townsend,
  arXiv:hep-th/0703276.
\bibitem{Kallosh}
  E.~Bergshoeff, R.~Kallosh, and T.~Ortin,
  Nucl.\ Phys.\  B {\bf 478}, 156 (1996).
\bibitem{Copeland}
  E.~J.~Copeland, R.~Easther, and D.~Wands,
  Phys.\ Rev.\  D {\bf 56}, 874 (1997).
\bibitem{Sugra}
  D.~Z.~Freedman and J.~H.~Schwarz,
  Nucl.\ Phys.\ B {\bf 137}, 333 (1978).
\bibitem{Sonner}
  J.~Sonner and P.~K.~Townsend,
  Phys.\ Rev.\  D {\bf 74}, 103508 (2006).
\bibitem{Dilaton}
  T.~Damour, G.~W.~Gibbons, and C.~Gundlach,
  Phys.\ Rev.\ Lett. {\bf 64}, 123 (1990);
  T.~Damour, F.~Piazza, and G.~Veneziano,
  Phys.\ Rev.\ Lett. {\bf 89}, 081601 (2002).
  M.~Gasperini,
  arXiv: hep-th/0702166.
\bibitem{Dilrun}
  M.~Gasperini, F.~Piazza, and G.~Veneziano,
  Phys.\ Rev.\ D {\bf 65}, 023508 (2002). 
\bibitem{CLW}
  E.~J.~Copeland, A.~R.~Liddle, and D.~Wands,
  Phys.\ Rev.\ D {\bf 57}, 4686 (1997).
\bibitem{Scaling}
  A.~J.~Liddle and R.~J.~Scherrer,
  Phys.\ Rev.\ D {\bf 59}, 023509 (1999);
  J.-P.~Uzan,
  Phys.\ Rev.\ D {\bf 59}, 123510 (1999);
  T.~Barreiro, E.~J.~Copeland, and N.~J.~Nunes,
  Phys.\ Rev.\ D {\bf 61}, 127301 (2000);  
  L.~Amendola, M.~Quartin, S.~Tsujikawa, and I.~Waga,
  Phys.\ Rev.\ D {\bf 74}, 023525 (2006).
\bibitem{Alam}
  U.~Alam, V.~Sahni, and A.~A.~Starobinsky,
  JCAP {\bf 0702}, 011 (2007).
\bibitem{STconstraints}
  A.~Riazuelo and J.~P.~Uzan,
  Phys.\ Rev.\  D {\bf 66}, 023525 (2002);
  B.~Bertotti, L.~Iess, and P.~Tortora,
  Nature, {\bf 425}, 374 (2003);
  T.~Damour and B.~Pichon,
  Phys.\ Rev.\  D {\bf 59}, 123502 (1999);
  A.~Coc, K.~A.~Olive, J.~P.~Uzan and E.~Vangioni,
  Phys.\ Rev.\  D {\bf 73}, 083525 (2006).
\bibitem{Wetterich}
  A.~Hebecker and C.~Wetterich,
  Phys.\ Lett.\ B {\bf 497} 281 (2001).
\end{thebibliography}
\end{document}